\documentclass[conference]{IEEEtran}
\IEEEoverridecommandlockouts
\usepackage{cite}
\usepackage{amsmath,amssymb,amsfonts}
\usepackage{algorithmic}
\usepackage{graphicx}
\usepackage{textcomp}
\usepackage{xcolor}
\def\BibTeX{{\rm B\kern-.05em{\sc i\kern-.025em b}\kern-.08em
    T\kern-.1667em\lower.7ex\hbox{E}\kern-.125emX}}

\usepackage{hyperref}
\usepackage{tikz}
\usetikzlibrary{positioning,chains,arrows,arrows.meta,shadows,fadings,shapes,backgrounds,matrix,patterns,plotmarks,trees,mindmap,automata,fit,calc,decorations.text}
\usepackage{xfp}
\usepackage{makecell}
\usepackage{booktabs}
\usepackage{listings}
\usepackage{color}
\usepackage[linesnumbered,ruled,vlined]{algorithm2e}
\usepackage{siunitx}
\usepackage{multirow}
\usepackage{cleveref}
\usepackage{mathcommands}
\usepackage{scheduling}

\newcommand{\tick}[0]{\text{tick}}

\begin{document}

\title{CHRONOS: Compensating Hardware Related Overheads with Native Multi Timer Support for Real-Time Operating Systems
    \thanks{This work has received funding from the DFG Priority Program ``Disruptive Memory Technologies'' (SPP 2377) as part of the project ``ARTS-NVM'' (502308721). It was further supported by the DFG Project ``One-Memory'' (405422836).}
}

\author{\IEEEauthorblockN{1\textsuperscript{st} Kay Heider}
    \IEEEauthorblockA{\textit{TU Dortmund University} \\
        Dortmund, Germany \\
        \href{mailto:kay.heider@tu-dortmund.de}{kay.heider@tu-dortmund.de}}
    \and
    \IEEEauthorblockN{2\textsuperscript{nd} Christian Hakert}
    \IEEEauthorblockA{\textit{TU Dortmund University} \\
        Dortmund, Germany \\
        \href{mailto:christian.hakert@tu-dortmund.de}{christian.hakert@tu-dortmund.de}}
    \and
    \IEEEauthorblockN{3\textsuperscript{rd} Kuan-Hsun Chen}
    \IEEEauthorblockA{\textit{University of Twente} \\
        Enschede, the Netherlands \\
        \href{mailto:k.h.chen@utwente.nl}{k.h.chen@utwente.nl}}
    \and
    \IEEEauthorblockN{4\textsuperscript{th} Jian-Jia Chen}
    \IEEEauthorblockA{\textit{TU Dortmund University} \\
        Dortmund, Germany \\
        \href{mailto:jian-jia.chen@cs.tu-dortmund.de}{jian-jia.chen@cs.tu-dortmund.de}}
}

\maketitle

\thispagestyle{plain}
\pagestyle{plain}

\begin{abstract}
    The management of timing constraints in a real-time operating system (RTOS) is usually realized through a global tick counter.
    This counter acts as the foundational time unit for all tasks in the systems.
    In order to establish a connection between a tick and an amount of elapsed time in the real world, often this tick counter is periodically incremented by a hardware timer.
    At a fixed interval, this timer generates an interrupt that increments the counter.
    In an RTOS, jobs can only become ready upon a timer tick.
    That means, during a tick interrupt, the tick counter will be incremented, jobs will be released, and potentially, a scheduling decision will be conducted to select a new job to be run.
    As this process naturally uses some processing time, it is beneficial regarding the system utilization to minimize the time spent in tick interrupts.
    In modern microcontrollers, multiple hardware timers are often available.
    To utilize multiple timers to reduce the overhead caused by tick interrupts, multiple methods are introduced in this paper.
    Generally, the task dispatching process is distributed over multiple timers, where each timer manages a subset of the task set.
    The number of interrupts that are triggered by these timers can then be reduced by mapping tasks to timers in such a manner that the greatest common divisor (GCD) of all task periods in a subset is maximized, and the GCD is adopted as the interrupt interval of the timer.
    To find an optimal mapping of tasks to timers, an MIQCP-model is presented that minimizes the overall number of tick interrupts that occur in a system, while ensuring a correct task release behavior.
    The presented methods are implemented in FreeRTOS and evaluated on an embedded system.
    The evaluation of the methods show, that compared to the baseline implementation in FreeRTOS that uses a single timer with a fixed period, the presented methods can provide a significant reduction in overhead of up to $\approx10\times$ in peak and up to $\approx 6\times$ in average.
\end{abstract}


\section{Introduction}\label{chapter:Introduction}
In a real-time operating system (RTOS), the notion of time is usually abstracted by a global system tick counter.
Often, this tick counter is periodically incremented by an underlying hardware timer that counts at a fixed frequency.
This system timer establishes the connection between the tick counter and the elapsed wall time.
Usually, the process of handling the timer tick is realized by an interrupt.

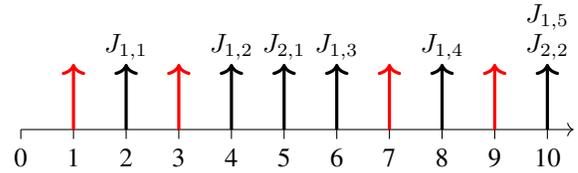
\begin{figure}
    \centering
    \begin{tikzpicture}[yscale=0.5, xscale=0.7]
        \timeline{0}{10.5}{}
        \labelling{0}{10}{1}{0}

        \releases{2,4,5,6,8,10}
        {\color{red} \releases{1,3,7,9}}
        \node at (2,2.2) {$J_{1,1}$};
        \node at (4,2.2) {$J_{1,2}$};
        \node at (5,2.2) {$J_{2,1}$};
        \node at (6,2.2) {$J_{1,3}$};
        \node at (8,2.2) {$J_{1,4}$};
        \node at (10,2.2) {$J_{2,2}$};
        \node at (10,3) {$J_{1,5}$};
    \end{tikzpicture}
    \caption{Example of tick interrupts with a single timer and the release times of jobs from two tasks $\tau_1$ and $\tau_2$. An upward arrow indicates a tick interrupt.}\label{fig:motivationt1}
\end{figure}

In addition to keeping the time in an RTOS, the tick counter is used to schedule and release jobs.
A job can only ever become ready upon a timer tick, and depending on the scheduling policy, that newly released job can be selected to be executed after the tick handling.
To correctly release jobs in a system with strictly periodic tasks, the tick interrupt needs to occur at a period, which is a common divisor of all task periods.
Otherwise, a job would miss the time it should become ready.
As a result, there are potentially tick interrupts where no job becomes ready, and only the tick counter is incremented.
For example, in a system where one task releases jobs every two time units and another task releases jobs every five time units, the system timer needs to tick at every time unit as both task periods are co-prime.
Suppose that both tasks release their first job at time zero. Then, the system execution is only affected by the interrupt that dispatches new jobs that at times $2,4,5,6,8,10,\dots$ and so forth.
The tick interrupt, however, occurs at every time unit.
In \Cref{fig:motivationt1}, the tick interrupts and release times of the jobs from these two tasks are depicted for a system with a single timer.
During the ten time units depicted, at four out of ten interrupts, no jobs become ready.
The process of incrementing the system tick counter, potentially releasing new jobs, and performing a scheduling decision, naturally causes an overhead for each handling of a timer tick.
To reduce this overhead, it is desirable to reduce the number of tick interrupts where no job becomes active, while maintaining a correct job release behavior.

Modern microprocessors often possess multiple hardware timers that can be configured to independently generate interrupts at different intervals.
For example, the ESP32 series of microprocessors contain four independent hardware timers~\cite{ESP32S3}.
An automotive microprocessor like the Infineon TC1796, that is, for example, used by the SLOTH on time, contains 256 hardware timers~\cite{Hofer2012}.
Utilizing multiple timers to manage a set of periodic real-time tasks can reduce the number of tick interrupts and thus reduce the overall overhead caused by tick interrupts.
This can be achieved by partitioning the task set into subsets where the greatest common divisor (GCD) of the task periods is maximized.
A separate hardware timer can then be assigned to manage only that partition of the task set, and thus, the tick interrupt interval can be set to the GCD of the task periods.
Considering the example from before, a single timer must tick at every time unit to ensure the correct dispatching of jobs.
For a system with at least two hardware timers, one timer could manage the task with period two, and another timer could manage the task with period five.
Thus, the tick interrupt intervals could also be set to two and five time units, respectively.
As a result, tick interrupts only occur when jobs are released, minimizing the overall number of tick interrupts.

\begin{figure}
    \centering
    \begin{tikzpicture}[yscale=0.5, xscale=0.7]
        \timeline{0}{10.5}{}
        \labelling{0}{10}{1}{0}

        \releases{2,4,5,6,8,10}
        \node at (2,2.2) {$J_{1,1}$};
        \node at (4,2.2) {$J_{1,2}$};
        \node at (5,2.2) {$J_{2,1}$};
        \node at (6,2.2) {$J_{1,3}$};
        \node at (8,2.2) {$J_{1,4}$};
        \node at (10,2.2) {$J_{2,2}$};
        \node at (10,3) {$J_{1,5}$};

        \node at (2,-1.5) {$\rho_1$};
        \node at (4,-1.5) {$\rho_1$};
        \node at (5,-1.5) {$\rho_2$};
        \node at (6,-1.5) {$\rho_1$};
        \node at (8,-1.5) {$\rho_1$};
        \node at (10,-2.2) {$\rho_2$};
        \node at (10,-1.5) {$\rho_1$};
    \end{tikzpicture}
    \caption{Example of tick interrupts with two timers $\rho_1,\rho_2$ and the release times of jobs from two tasks $\tau_1,\tau_2$. An upward arrow indicates a tick interrupt.}\label{fig:motivationt2}
\end{figure}
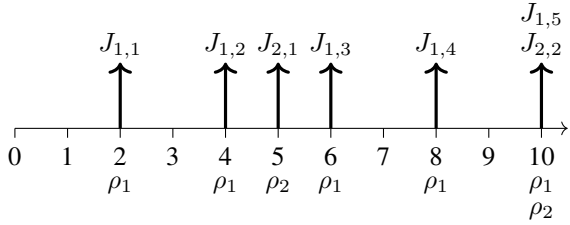

In \Cref{fig:motivationt2}, the tick interrupts and release times of the two jobs are depicted for a system with two timers $\rho_1,\rho_2$.
During the depicted ten time units, a job becomes ready at every tick interrupt.
Since the timers are configured such that they trigger an interrupt only at the time when a job will be released, there are no interrupts that cause unnecessary overhead.
It should be noted that at time ten, two interrupts are triggered, one from each timer.

In this paper, we examine strategies to reduce the task handling overhead incurred by housekeeping operations of an RTOS for strictly periodic task sets.
As shown by a recent empirical industry study, in 82\% of the investigated systems, task activations are periodic~\cite{Akesson2020}.
Towards optimizing the task handling process, we present an MIQCP that is used to partition a given task set and configure separate hardware timers to handle those partitions to minimize the overall number of tick interrupts.
Moreover, we examine different strategies to manage the set of delayed tasks which has an impact on the efficiency of delaying tasks and releasing jobs.

\section{Related Work}\label{Related Work}
Many methods to improve the system efficiency of real-time operating systems (RTOSes) have been proposed before.
Hofer et~al. propose a method where, for every task, multiple hardware timer cells are used, among other things, to control the activation of the task and to monitor the deadline of a task~\cite{Hofer2012}.
Their approach achieves a very low task dispatching latency, resulting in an improvement of up to $171\times$.
To improve the efficiency of managing timers in a system, Varghese and Lauck discuss several implementation schemes and propose the use of a timing wheel which effectively realizes a bucket sorting technique for systems where there is enough memory available~\cite{Varghese1997}. Their approach can perform management operations of timers in constant time.
Another approach to realize efficient timers is given by Aron and Druschel~\cite{softtimer}.
They propose soft timers which are triggered at certain times during the execution of the system where the overhead of invoking a timer handler is very low.
In order to more efficiently realize typical operations in an RTOS, such as task scheduling and time management, Kohout et~al. propose a hardware module that realizes these operations~\cite{Kohout}.
Their approach decreases the processing time for these RTOS operations by up to $90\%$.
Hagens and Chen investigate the efficiency of using different data structures as a basis for the task dispatcher in FreeRTOS, an open-source RTOS that will also be used in this paper~\cite{hagens}.
They evaluate the usage of Lists, Binary Search Trees, Red-Black Trees, and Heaps.
Moreover, Balas and Benini also consider FreeRTOS, but on a RISC-V-based microcontroller~\cite{9474114}.
They utilize specific instructions of the RISC-V instruction set to improve the handling of interrupts and to reduce the time needed for context switches.

\section{System Model}\label{System Model}
We consider an RTOS with a periodic timer tick interrupt that invokes a \textit{system tick handler}.
This interrupt handler is responsible for central housekeeping operations of the RTOS, namely incrementing the global tick counter, releasing new jobs, and conducting a scheduling decision whenever a new job is released.
Therefore, new jobs are solely released during the system tick interrupt.
Generally, the system tick handler can be invoked by a timer that generates interrupts at a fixed interval, i.e., every $P$ time units, the interrupt will occur, or by a \textit{one-shot} timer whose alarm interval is reconfigured after every occurrence.
In this paper, we consider RTOSes where the system tick is backed by a timer interrupt with a fixed interrupt interval during runtime.

Moreover, we consider task sets that are comprised of strictly periodic tasks that release their first jobs synchronously at time 0.
Hence, any job is always released at time points that are integer multiples of their corresponding task's period.
We denote $\mathbb{T}=\{\tau_1,\dots,\tau_n\}$ as the task set of $n$ tasks that should be executed on the system.
A periodic task is denoted by $\tau_i=(C_i,T_i,D_i)$ with $C_i$ as the worst-case execution time (WCET), $T_i$ the period of the task, and $D_i$ the relative deadline.
Every task releases infinitely many jobs, the set of jobs released by task $\tau_i$ are denoted as $\mathbb{J}_i$, and the $k$-th released job of $\tau_i$ is denoted as $J_{i,k}$.
Since every task releases their first job at time 0, the release time of the $k$-th job of task $i$ is given by $k\cdot T_i$.
Additionally, we assume that the task set, along with the number of timers and the mapping of tasks to timers, are given.
The number of timers and the mapping are also assumed to be fixed during runtime.

Regarding the hardware side, we assume a CPU with multiple hardware timers, whose interrupt intervals can be independently chosen.
The interrupt intervals are assumed to be configurable as an integer multiple of a base time interval that is specified by the hardware.
After a timer has triggered an interrupt, the internal counter of the hardware timer should be reset to zero.
Alternatively, a timer could trigger interrupts whenever the internal counter has reached an integer multiple of a given time interval.
The set of configurable timers is denoted as $\mathbb{P}=\{\rho_1,\dots,\rho_m\}$, with timer $\rho_j$ triggering an interrupt every $P_j$ time units.

\section{Multi-Timer Task Dispatching}\label{section:Multi-Timer}
Generally, for a synchronous task set that contains only periodic tasks, the release times of the tasks are always at tick counts that are multiples of the task periods.
To ensure a correct task release behavior, the tick interrupt needs to occur at an interval that is a divisor of all task periods.
Consequently, if any two task periods are co-prime, the tick interrupt needs to occur at every time unit.
However, if the GCD of all task periods is greater than one, it is possible to trigger an interrupt at an interval given by the GCD of the task periods.
The ticks that are required to ensure a correct task release behavior, i.e., ticks where tasks are released, are called \textit{required} ticks.
A tick where no task can be released is called a \textit{not-required} tick.
It is desirable, regarding the improvement of the system efficiency, that the number of \textit{not-required} ticks is reduced, as every interrupt causes a certain overhead.
The notion of \textit{required} and \textit{not-required} ticks can be explained by \Cref{fig:motivationt1} using the example from the introduction.
Here, a red upward arrow indicates a \textit{not-required} tick and a black upward arrow indicates a \textit{required} tick.
As the GCD of the task periods is one, the timer period $P_j$ must also be set to one.
In this scenario, the ticks $1,3,7$, and $9$ are \textit{not-required} as, there, no task becomes ready.

When a system can harness multiple hardware timers, it is possible to distribute the task handling process over multiple timers.
Instead of using a single timer to manage the releases of all tasks, every hardware timer can manage a subset of tasks where the GCD of the task periods is potentially greater.
The GCD in that subset can then be adopted as the tick interrupt interval of the associated hardware timer.
The idea is shown in Figure~\ref{fig:distribution} and described in more detail in the following:

\begin{figure}
    \centering
    \begin{tikzpicture}
        [rect/.style={ rectangle,
                    thick,
                    shape aspect=2,
                    inner sep = 5pt,
                    text centered,
                    minimum width = 1cm,
                    minimum height = .4cm,
                }]
        \node[rect,draw,anchor=north west,minimum height=4.8cm,minimum width=1cm] (taskset) at (0,0) {};
        \node[rect,draw,anchor=north west,minimum height=1cm,minimum width=1cm] (h0) at (5,-0.1) {$\rho_1$};
        \node[rect,draw,anchor=north west,minimum height=1cm,minimum width=1cm] (h1) at (5,-1.3) {$\rho_2$};
        \node[rect,anchor=north west,minimum height=1cm,minimum width=1cm] (hd) at (5,-2.4) {\vdots};
        \node[rect,draw,anchor=north west,minimum height=1cm,minimum width=1cm] (hn) at (5,-3.7) {$\rho_m$};

        \node at (0.5,-0.6) {$\mathbb{T}_1$};
        \node at (0.5,-1.8) {$\mathbb{T}_2$};
        \node at (0.5,-2.9) {$\vdots$};
        \node at (0.5,-4.2) {$\mathbb{T}_k$};

        \node at ([yshift=2ex]taskset.north) {Task Set $\mathbb{T}$};
        \node at ([yshift=2ex]h0.north) {Timers};

        \draw[thick] (0,-1.2) -- (1,-1.2);

        \draw[thick] (0,-2.4) -- (1,-2.4);

        \draw[thick] (0,-3.6) -- (1,-3.6);

        \path[-Latex,thick] (taskset.east|- h0.west) edge node [above] {\small $P_1=\text{GCD}(\mathbb{T}_1)$} (h0);
        \path[-Latex,thick] (taskset.east|- h1.west) edge node [above] {\small $P_2=\text{GCD}(\mathbb{T}_2)$} (h1);
        \path[-Latex,thick] (taskset.east|- hn.west) edge node [above] {\small $P_m=\text{GCD}(\mathbb{T}_k)$} (hn);

    \end{tikzpicture}
    \caption{Mapping of $k$ partitions of a task set $\mathbb{T}$ to $m$ timers}\label{fig:distribution}
\end{figure}
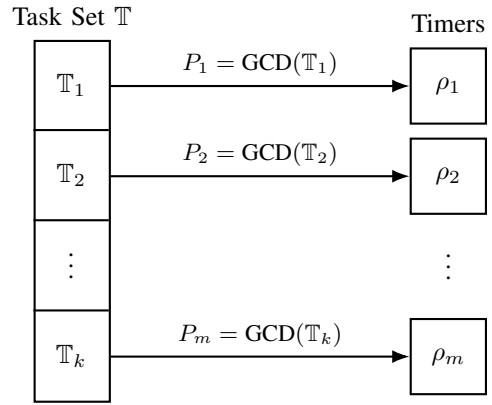

For a task set $\mathbb{T}$ that contains $n$ periodic tasks, there are at most $n$ different task periods.
When employing multiple hardware timers $\rho_1,\dots,\rho_m$ to distribute the task handling process over, every timer $\rho_j$ can be assigned a subset $\mathbb{T}_j\subseteq\mathbb{T}$.
In the extreme case, every hardware timer could be assigned a subset in which the task period is uniform, i.e., the system would need at most $n$ hardware timers such that no tick interrupts occur for \textit{not-required} ticks.
However, this is unrealistic, as in most systems, there are significantly fewer hardware timers than tasks with different periods.
A more sensible approach is to partition the task set into $k$ subsets $\mathbb{T}_1,\dots,\mathbb{T}_k$ that cover multiple tasks.
A timer $\rho_j$ is then configured to maintain the tasks of the set $\mathbb{T}_j$, for $1\leq j\leq k$.
Intuitively, to reduce the number of tick interrupts, the task set should be partitioned such that the GCD of the task periods of each partition is maximized.
However, more attention needs to be paid to the number of timers employed and their configured periods:

Suppose that the system contains $m$ hardware timers $\rho_1,\dots,\rho_m$ that trigger periodic interrupts at $P_{1},\dots,P_{m}$ time units.
The expected number of tick interrupts that occur for a fixed time interval in the multi-timer system can be calculated as:
\begin{equation}
    \sum_{j=1}^{m}\frac{1}{P_j}\label{ex timer}
\end{equation}

As a result, the overall number of tick interrupts is generally reduced when the timer periods are increased.
However, special attention has to be given to situations where it is not always beneficial to utilize multiple timers even though they tick at longer intervals than it would be possible with just a single timer.
For example, suppose that the task set consists of three tasks with periods $T_1=2$, $T_2=3$, and $T_3=5$.
In the system with a single timer, a tick interrupt needs to occur at every time unit as the task periods are co-prime, while for a system with three hardware timers $\rho_1,\rho_2,\rho_3$ their interrupt intervals could be set to $P_1=2$, $P_2=3$, and $P_3=5$ time units.
The system with multiple timers then experiences $\frac{1}{2}+\frac{1}{3}+\frac{1}{5}>1$ tick interrupts under the observation from~\Cref{ex timer}, i.e., more interrupts will be triggered compared to the single timer system, and thus more overhead is caused.

A further observation is that there is naturally no benefit in having multiple timers whose interrupt intervals are integer multiples of each other.
Specifically, if the GCD in any subset was equal to one, the associated timer would need to interrupt at every time unit and there is no benefit to employ any additional timers.

Naturally, it is not always possible to find a partition that assigns multiple tasks to each timer where overall the number of tick interrupts are reduced.
Nonetheless, if the system contains a sufficient number of hardware timers for a task set, the overall overhead can potentially be reduced.
A method to find an optimal partition of the task set that minimizes the number of tick interrupts is presented in Section~\ref{section:Task Distribution}.

\section{Chronos}\label{sec:Chronos}
In order to exploit the observations made in Section~\ref{section:Multi-Timer}, the task handling process has to be distributed over multiple timers.
We present \textit{Chronos}, a method that uses multiple timers with different interrupt intervals to release jobs from a given partition of a task set.
The general idea is to create multiple tick counters and multiple lists to store delayed tasks that are each managed by a separate hardware timer.
Each timer then maintains the task release process for a partition of the task set.
We assume that the set of delayed tasks is kept sorted.

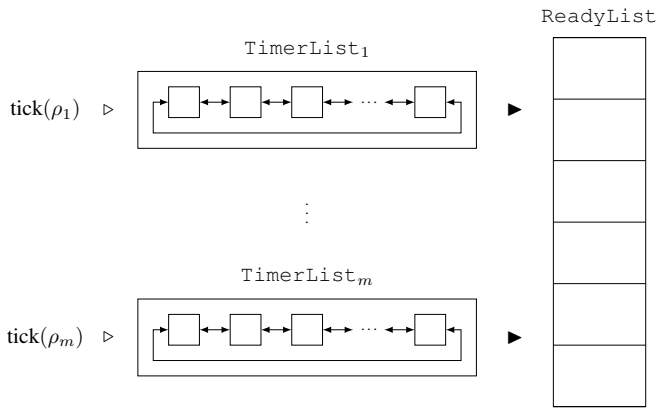
\begin{figure}
    \centering
    \resizebox{\linewidth}{!}{%
        \begin{tikzpicture}
            \foreach \y in {-2,1.7}{
                    \foreach [count=\i] \s in {0.25,1.25,2.25,4.25} {
                            \draw (\s,\y+0) rectangle  (\s + .5,\y+.5) node[pos=.5] {};
                        }
                    \draw[draw=white] (3.25,\y+0) rectangle  (3.5+.25,\y+.5) node[pos=.5,xshift=.75] {\scriptsize\dots};
                    \foreach [count=\i] \s in {.75,1.75,2.75,3.75} {
                            \draw[latex-latex] (\s,\y+0.25) -- (\s + .5,\y+.25);
                        }
                    \draw[latex-latex] (4.75,\y+0.25) -- (5,\y+0.25) -- (5,\y+-.25) -- (0,\y+-.25) -- (0,\y+.25)-- (.25,\y+.25);
                };
            \draw[draw=white] (-2.5,1.7+-.5) rectangle (-1,1.7+.75) node[pos=0.5] {$\tick(\rho_1)$};
            \draw[draw=white] (-2.5,-2+-.5) rectangle (-1,-2+.75) node[pos=0.5] {$\tick(\rho_m)$};

            \draw[draw=white] (-1.2+\fpeval{0.5/2},\fpeval{1.2+((1.7+0.75)-(1.7+-0.5))/2}) rectangle (-.25-\fpeval{0.5/2},\fpeval{1.2+((1.7+0.75)-(1.7+-0.5))/2}) node[pos=0.5] {$\triangleright$};
            \draw[draw=white] (-1.2+\fpeval{0.5/2},\fpeval{-2.5+((-2+.75)-(-2+-.5))/2}) rectangle (-.25-\fpeval{0.5/2},\fpeval{-2.5+((-2+.75)-(-2+-.5))/2}) node[pos=0.5] {$\triangleright$};

            \draw (-.25,-2+-.5) rectangle  (5.25,-2+.75) node[pos=.5,yshift=1cm] {$\texttt{TimerList}_m$};

            \draw[draw=white] (2.25,0+0) rectangle  (2.25+.5,0+.5) node[pos=.5] {\scriptsize\vdots};

            \draw (-.25,1.7+-.5) rectangle  (5.25,1.7+.75) node[pos=.5,yshift=1cm] {$\texttt{TimerList}_1$};

            \draw (6.5,-3) rectangle (8,3) node[pos=.5,yshift=3.35cm] {\texttt{ReadyList}};
            \foreach [count=\i] \s in {0,1,2,3,4,5,6} {
                    \draw (6.5,-3+\s) -- (8,-3+\s);
                }
            \draw[draw=white] (5.375,\fpeval{1.2+((1.7+0.75)-(1.7+-0.5))/2}) rectangle (6.375,\fpeval{1.2+((1.7+0.75)-(1.7+-0.5))/2}) node[pos=0.5] {$\blacktriangleright$};
            \draw[draw=white] (5.375,\fpeval{-2.5+((-2+.75)-(-2+-.5))/2}) rectangle (6.375,\fpeval{-2.5+((-2+.75)-(-2+-.5))/2})node[pos=0.5] {$\blacktriangleright$};
        \end{tikzpicture}}
    \caption{Dispatcher architecture with multiple timers}\label{fig:Chronos}
\end{figure}

\IncMargin{1.5em}%
\begin{algorithm}
    \SetAlgoLined
    \SetKwInput{Input}{Input}
    \SetKwInput{Assume}{Require}
    \SetKw{Break}{break}
    \SetKwFor{Loop}{loop}{}{end loop}
    \DontPrintSemicolon

    \Indentp{-1.5em}
    \Input{Timer number $j$}
    \Assume{$P_j>0$, $\texttt{TimerList}_j$ sorted ascending by tasks' \texttt{nextRelease}}
    \Indentp{1.5em}
    \BlankLine
    $\tick(\rho_j)$ += $P_j$\;
    \If {$\emph{\tick}(\rho_j)\geq\emph{\texttt{timerNextRelease}}_j$}{
        \Loop{}{
            \If {$\emph{\texttt{TimerList}}_j$ is empty}{
                $\texttt{timerNextRelease}_j=\texttt{MAX\_VALUE}$\;
                \Break
            }
            $\tau\gets head(\texttt{TimerList}_j)$\;
            \If {$\tau\emph{\texttt{.nextRelease}} > tick(\rho_j)$}{
                $\texttt{timerNextRelease}_j=\tau\texttt{.nextRelease}$\;
                \Break
            }
            remove $\tau$ from $\texttt{TimerList}_j$\;
            add $\tau$ to \texttt{ReadyList}\;
        }
    }
    \caption{\textit{Chronos} Interrupt Routine}\label{Chronos}
\end{algorithm}\DecMargin{1.5em}

\paragraph{Architecture}\label{sec:Chronos arch}
\textit{Chronos} naturally extends the concept of a single timer that manages a global tick and a global list for delayed tasks to multiple timer sources.
Given a system that contains $m$ hardware timers, the mapping of tasks to timers, that is supplied as an input, induces a partition of the task set into at most $m$ subsets.
Each subset $\mathbb{T}_j$ contains the tasks that shall be managed by timer $\rho_j$.
The period $P_j$ of timer $\rho_j$ is configured to the GCD of the periods of all tasks in $\mathbb{T}_j$.
If all available $m$ timers are used, $m$ tick counters and $m$ lists for delayed tasks will be kept.
Each timer $\rho_j$ independently manages a tick counter $\tick(\rho_j)$ and a list $\texttt{TimerList}_j$ that is used to store delayed tasks.
A schematic for the architecture can be found in Figure~\ref{fig:Chronos}.

Instead of a global tick count that is used by all tasks as a time reference, an independent tick counter is created for each hardware timer.
Each of these counters is incremented whenever the associated timer triggers an interrupt.
To still have a unified length for a tick over all timers, the tick counters are incremented by the period of the timer.
It is assumed that all timers are configured such that the basic time unit for all timers is the same, e.g., all timers generate interrupts at multiples of $\SI{1}{\milli\second}$ such that the time unit of task and timer periods is also given in \SI{}{\milli\second}.
Tasks that are assigned to a specific timer then use the timer's tick counter as a reference.

Alongside a tick counter $\tick(\rho_j)$, a list $\texttt{TimerList}_j$ is created for each hardware timer $\rho_j$.
This $\texttt{TimerList}_j$ is used to store delayed tasks that are assigned to the timer $\rho_j$.
We assume that the tasks are inserted into the $\texttt{TimerList}_j$ sorted by their next release time in ascending order.
Compared to a system with one global timer, the number of tasks that are managed by a single timer in \textit{Chronos} is usually lower as the tasks are distributed over multiple timers.
This is beneficial in terms of the overhead for inserting tasks.

Furthermore, a function \texttt{delay()} is introduced delays a periodic task in a multi-timer setup.
This function computes the next time when a job of the task should be released and inserts the delayed task to the $\texttt{TimerList}_j$ of the associated timer $\rho_j$ in a sorted manner.
The list is sorted by the next release times in ascending order.
Also, a $\texttt{timerNextRelease}_j$ value is kept for each timer $\rho_j$ that is always set to the next tick count when a task assigned to $\rho_j$ should become ready.

\paragraph{Interrupt Routine}
As the number of timers that will be used by \textit{Chronos} is variable, the interrupt routine for the hardware timers is not specifically designed for each timer but bases its functionality on the timer that executes it.
Algorithm~\ref{Chronos} shows pseudocode of the interrupt routine.

The \textit{Chronos} tick interrupt routine is executed by all hardware timers and takes as input an identifier of the timer that triggers the input, i.e., if timer $\rho_j$ triggers the interrupt, the index $j$ will be used to determine which tick counter and which $\texttt{TimerList}_j$ has to be updated.
Generally, it is assumed that the timer period $P_j$ is greater than zero and that the associated $\texttt{TimerList}_j$ is sorted by the task's next release time in ascending order.

First, the tick counter $\tick(\rho_j)$ is incremented by the timers period $P_j$.
After this, it is checked whether a task can become ready by comparing the updated tick counter $\tick(\rho_j)$ with the $\texttt{timerNextRelease}_j$ value.
If it is not time to release a task yet, the interrupt exits early.
Conversely, if the tick count has reached the time when a task can be released, the associated $\texttt{TimerList}_j$ is iterated.
In the case that the $\texttt{TimerList}_j$ becomes empty because all tasks were released, the $\texttt{timerNextRelease}_j$ is set to a maximal value, such that if multiple interrupts are triggered without any task being added to the list in between, the interrupt routine can always exit early.
Otherwise, the task that is at the head of the $\texttt{TimerList}_j$ will be inspected.
If the next release time of the inspected task $\tau$ has not been reached yet, the $\texttt{timerNextRelease}_j$ is set to the next release time of that task, and the routine exits.
If the task $\tau$ should be released, it is removed from the $\texttt{TimerList}_j$ and added to the global \texttt{ReadyList}.
This is done until a task is encountered that cannot be released yet or the $\texttt{TimerList}_j$ is empty.

\subsection{Chronos-const}\label{sec:Chronos-const}
As \textit{Chronos} aims to reduce the number of tick interrupts, it can be explored whether it is beneficial to reduce the overhead of inserting tasks into a $\texttt{TimerList}_j$ in a sorted manner at the cost of increasing the workload in each tick interrupt.
The method called \textit{Chronos-const} realizes a constant cost for inserting tasks into a $\texttt{TimerList}_j$ with the trade-off that a tick interrupt now always causes a linear overhead regarding the number of tasks contained in the $\texttt{TimerList}_j$.

Generally, \textit{Chronos-const} uses the same architecture as \textit{Chronos}; however, the way each $\texttt{TimerList}_j$ is managed changes.
To achieve a constant cost for inserting tasks into a $\texttt{TimerList}_j$, tasks will now be appended to the list.
That means there is no order that can be exploited in the tick interrupt.
To determine which tasks should be released, the whole $\texttt{TimerList}_j$ has to be inspected during an interrupt of timer $\rho_j$.
For every task, its next release time will be examined, and the task will be released accordingly.
This causes a linear overhead regarding the number of delayed tasks contained in the $\texttt{TimerList}_j$ at the point when the interrupt is triggered.
The tick interrupt routine is stated in pseudocode in Algorithm~\ref{Chronos-const}.

\IncMargin{1.5em}%
\begin{algorithm}
    \SetAlgoLined
    \SetKwInput{Input}{Input}
    \SetKwInput{Assume}{Require}
    \SetKw{Break}{break}
    \SetKwFor{Loop}{loop}{}{end loop}
    \DontPrintSemicolon

    \Indentp{-1.5em}
    \Input{Timer number $j$}
    \Assume{$P_j>0$}
    \Indentp{1.5em}
    \BlankLine
    $\tick(\rho_j)$ += $P_j$\;
    \If {$\emph{\tick}(\rho_j)\geq\emph{\texttt{timerNextRelease}}_j$}{
        $\texttt{timerNextRelease}_j=\texttt{MAX\_VALUE}$\;
        $\tau\gets head(\texttt{TimerList}_j)$\;
        \While{$\tau\neq tail(\emph{\texttt{TimerList}}_j)$}{
            \eIf {$\tau.\emph{\texttt{nextRelease}} > \emph{\textrm{tick}}(\rho_j)$}{
                \If {$\tau.\emph{\texttt{nextRelease}} < \emph{\texttt{timerNextRelease}}_j$}{
                    $\texttt{timerNextRelease}_j=\tau\texttt{.nextRelease}$\;
                }
                $\tau\gets next(\texttt{TimerList}_j)$\;
            }{
                $\tau'\gets next(\texttt{TimerList}_j)$\;
                remove $\tau$ from $\texttt{TimerList}_j$\;
                add $\tau$ to \texttt{ReadyList}\;
                $\tau\gets \tau'$\;
            }
        }
    }
    \caption{\textit{Chronos-const} Interrupt Routine}\label{Chronos-const}
\end{algorithm}\DecMargin{1.5em}

Like the \textit{Chronos} tick interrupt routine, the \textit{Chronos-const} tick interrupt routine is executed by all hardware timers and determines which timer structures should be updated on the basis of the supplied timer number.
Here, it is also assumed that timer periods are strictly greater than zero, but the $\texttt{TimerList}_j$ does not need to be sorted.

The first step in the interrupt routine is to increment the tick counter $\tick(\rho_j)$ by the timer's period $P_j$.
If the tick count has not reached the time when a task can be released, the routine exits early.
Otherwise, initially, the $\texttt{timerNextRelease}_j$ will be set to a maximal value.
Over the course of the interrupt, the $\texttt{timerNextRelease}_j$ will always be updated if a task is encountered whose next release time is lower than the current $\texttt{timerNextRelease}_j$.
That means, at the end of the routine, the $\texttt{timerNextRelease}_j$ will be set to the minimal next release time of any contained task.
The $\texttt{TimerList}_j$ will always be iterated completely, i.e., the routine exists only after the tail of the $\texttt{TimerList}_j$ is reached.
The tail of a list is assumed to be defined to be a special item that marks the end of the list.
If the currently examined task cannot be released, the $\texttt{timerNextRelease}_j$ will be adjusted accordingly, and the iteration continues with the next task.
However, if the task can be released, first, the task's successor in the list will be stored in $\tau'$.
This is because if $\tau$ is removed from the list, the current position in the list would be lost.
Afterwards, $\tau$ is removed from the $\texttt{TimerList}_j$ and inserted into the global \texttt{ReadyList}.
Finally, the iteration continues by setting $\tau$ to $\tau'$, restoring the position in the list and effectively advancing by one item.

\subsection{Chronos-harmonic}\label{sec:Chronos-harmonic}
The first two presented methods are applicable to any task set comprised of only periodic tasks and can result in a reduced number of tick interrupts if a suitable partition is found.
For harmonic task sets, additional optimizations can be made.
\textit{Chronos-harmonic} exploits the fact that the order in which tasks from a harmonic task set are released always stays the same.
In fact, the release pattern is given by the periods of the tasks.
Generally, \textit{Chronos-harmonic} can be applied to harmonic task sets where all tasks are periodic and released at time zero.
Additionally, \textit{Chronos-harmonic} can be applied to task sets that are not harmonic as a whole, but where a harmonic base exists that contains at most as many elements as there are available hardware timers.
Then, the tasks with periods of each harmonic chain can be managed by a separate hardware timer.
With \textit{Chronos-harmonic}, the cost of delaying a task is constant without increasing the overhead of the tick interrupt.

Since this method is specifically designed for harmonic task sets, it should be discussed whether harmonic task sets are in fact encountered in real systems.
For example, in many avionic systems, the task periods are harmonic, as described by Easwaran et~al.~\cite{Easwaran2009}.
Harmonic task sets are also considered for applications in the field of robotics~\cite{Li2003}, while in automotive systems or submarine systems, many tasks also have harmonic periods~\cite{Kramer2015,BusquetsMataix1996}.
Therefore, much research also focuses on harmonic task sets~\cite{Kuo1991,Nasri2014,Brueggen2017}.

\paragraph{Architecture}\label{sec: Chronos-harmonic arch}
The architecture of \textit{Chronos-harmonic} is still similar to \textit{Chronos} in terms of the multi-timer capabilities, as it distributes the task dispatching process over multiple hardware timers by introducing local structures for every timer.
That means, for each timer $\rho_j$, a $\texttt{TimerList}_j$ and a tick counter $\tick(\rho_j)$ is kept.
However, here, the $\texttt{TimerList}_j$ is implemented as an array of fixed-size that contains all tasks a timer manages, sorted after their period in ascending order.
In more detail, for a timer $\rho_j$ and the associated harmonic task set $\mathbb{T}_j=\{\tau_1,\dots,\tau_k\}\subseteq\mathbb{T}$ with $1\leq k\leq \lvert\mathbb{T}\rvert$, the $\texttt{TimerList}_j$ is an array of $k$ elements that stores the tasks $\tau_1,\dots,\tau_k$ sorted ascending by their periods $T_1,\dots,T_k$.
As the task set is given and no tasks are allowed to be created during runtime, the task set $\mathbb{T}_j$ can be sorted offline before the scheduler is started.
Since $\mathbb{T}_j$ is harmonic, for any two task $\tau_1,\tau_2\in\mathbb{T}_j$ with $T_1\geq T_2$, $T_1$ is an integer multiple of $T_2$.
Consequently, whenever $T_1$ is released, $T_2$ and any other task in $\mathbb{T}_j$ with a smaller period than $T_1$ will also be released.
In \textit{Chronos-harmonic}, the period $P_j$ of the associated timer $\rho_j$ is configured to the smallest period of any task in $\mathbb{T}_j$, i.e., $P_j=\min\{T_i\mid \tau_i\in\mathbb{T}_j\}$.
Therefore, every time $\rho_j$ triggers an interrupt, at least one task will be released, i.e., the task with period $P_j$.
Also, if $\tick(\rho_j)\mod T_l=0$, for any task period $T_l\in \{T_i\mid 1\leq i\leq k\}$, all tasks in $\mathbb{T}_j$ with periods that are smaller than $T_l$ have to be released.
That means the order in which tasks are released is always fixed and determined by the tasks' periods.
As a result, there is no need to dynamically insert and sort the $\texttt{TimerList}_j$.

\paragraph{Interrupt Routine}
In \textit{Chronos-harmonic}, this observation is exploited by the fixed-sized array that contains the tasks in combination with an array that contains the task periods in the same order.
That means, for a task set with $n$ tasks, the index $i$, with $1\leq i\leq n$, selects the task $\tau_i$ in the $\texttt{TimerList}_j$, and the tasks period $T_i$ in the $\texttt{TaskPeriods}_j$ array.
Generally, whenever a task $\tau_i$ is released, it will be removed from the $\texttt{TimerList}_j$ by setting the field $\texttt{TimerList}_j\texttt{[i]}$ to \texttt{NULL}.
This marks that the task $\tau_i$ is currently ready and can be used to detect a potential deadline miss if the task is not in the $\texttt{TimerList}_j$ when an interrupt is triggered where $\tau_i$ should be released.

Now, in every interrupt for timer $\rho_j$, the $\texttt{TaskPeriods}_j$ array is iterated from the start until a task period $T_l$ is encountered where $\tick(\rho_j)\mod T_l\neq 0$ or the end of the array has been reached.
As the periods are sorted in ascending order, once a task has been encountered that cannot be released, further tasks will not be released either.
When a task $\tau_i\in\mathbb{T}_j$ should be released, $\tau_i$ has to be in the $\texttt{TimerList}_j$, i.e., the entry $\texttt{TimerList}_j\texttt{[i]}$ cannot be \texttt{NULL}.
If the entry is \texttt{NULL}, the task is skipped, and the next task is examined.
When a task $\tau_i$ is released, the entry $\texttt{TimerList}_j\texttt{[i]}$ is set to \texttt{NULL}, and $\tau_i$ is inserted into the global ready queue.
The described tick interrupt routine is given as pseudocode in Algorithm~\ref{Chronos-harmonic}.

\IncMargin{1.5em}%
\begin{algorithm}
    \SetAlgoLined
    \SetKwInput{Input}{Input}
    \SetKwInput{Assume}{Require}
    \SetKw{Break}{break}
    \SetKw{Continue}{continue}
    \SetKwFor{Loop}{loop}{}{end loop}
    \DontPrintSemicolon

    \Indentp{-1.5em}
    \Input{Timer number $j$}
    \Assume{$P_j>0$, $\texttt{TimerList}_j$ sorted ascending by task periods}
    \Indentp{1.5em}
    \BlankLine
    $\tick(\rho_j)$ += $P_j$\;
    \For{$i\gets 1$ \KwTo $\lvert\mathbb{T}_j\rvert$}{
        \If{$\emph{\tick}(\rho_j)\mod T_i\neq 0$}{
            \Break
        }
        \If{$\emph{\texttt{TimerList}}_j\emph{\texttt{[i]}}=\emph{\texttt{NULL}}$}{
            \Continue
        }
        $\tau_i\gets \texttt{TimerList}_j\texttt{[i]}$\;
        $\texttt{TimerList}_j\texttt{[i]} = \texttt{NULL}$\;
        add $\tau_i$ to \texttt{ReadyList}\;
    }
    \caption{\textit{Chronos-harmonic} Interrupt Routine}\label{Chronos-harmonic}
\end{algorithm}\DecMargin{1.5em}

\section{Mapping Tasks to Timers}\label{section:Task Distribution}
In order to utilize the methods described in Section~\ref{sec:Chronos}, a mapping of tasks to timers is needed.
This mapping is critical to the efficiency of the methods, as it directly influences the number of tick interrupts that occur.
For this reason, it is desirable to find an optimal mapping such that the overall number of tick interrupts is minimized.

\subsection{Problem Definition}\label{sec: ilp problem}
As described in \Cref{section:Multi-Timer}, the number of tick interrupts that occur for not-required ticks should be minimized.
This works by mapping each task to a timer, inducing the partition of task set into $m$ subsets, for $m$ timers.
Concretely, every task in the same subset of the partition $\Gamma=\{\mathbb{T}_1,\dots,\mathbb{T}_m\}$ of the task set $\mathbb{T}=\{\tau_1,\dots,\tau_n\}$ will be assigned to the same timer $\rho_j$ for $1\leq j\leq m$.
The goal is then to minimize the expected number of tick interrupts.
This objective can be written as:

\begin{equation}
    \min\sum^{m}_{j=1}\frac{1}{P_j}\label{ilp: objective function}
\end{equation}

for $m$ timers with timer periods $P_1,\dots,P_m$.
It should be noted that the system could possess more than $m$ timers, as it is not always optimal to utilize all available timers.
For a timer $\rho_j$ to correctly release the tasks of its associated task set $\mathbb{T}_j$, the timer period $P_j$ must be an integer divisor of all task periods in $\mathcal{T}_j=\{T_i\mid\tau_i\in\mathbb{T}_j\}$.
It is assumed, without loss of generality, that any two tasks $\tau_1,\tau_2\in\mathbb{T}$ have different task periods $T_1,T_2$, since tasks with the same period can be assigned to the same timer $\rho_j$ without impacting $P_j$.
In order to minimize $\frac{1}{P_j}$ for any timer $\rho_j$, the timer period $P_j$ has to be maximized, i.e., $P_j$ should be set to the GCD of $\mathcal{T}_j$.

\subsection{Minimizing the Number of Tick Interrupts}\label{sec: miqcp}
\renewcommand{\theequation}{\thesection.\arabic{equation}}
\setcounter{equation}{0}
The problem described in Section~\ref{sec: ilp problem} can intuitively be described as an MIQCP model.

Let $\mathbb{T}=\{\tau_1,\dots,\tau_n\}$ be set of tasks with periods $\mathcal{T}=\{T_i\mid 1\leq i\leq n\}$.
Additionally, let $m\in\mathbb{N}_{\geq1}$ be the number of timers that are available in the system.
Then, the problem can be formulated as the following MIQCP model:
\begin{align}
    \min\quad         & \sum\limits^m_{j=1}f_j\cdot u_j\label{miqcp obj}                                                            \\
    \text{s.t.} \quad & f_j\cdot P_j = 1                                          &  & \text{for } j=1,\dots,m\label{miqcp num}     \\
                      & \sum\limits^m_{j=1}m_{i,j} = 1                            &  & \text{for } i=1,\dots,n\label{miqcp once}    \\
                      & d_i\cdot\sum\limits^m_{j=1} P_j\cdot m_{i,j} = T_i        &  & \text{for } i=1,\dots,n\label{miqcp divisor} \\
                      & \bigvee\limits^n_{i=1}m_{i,j} = u_j                       &  & \text{for } j=1,\dots,m\label{miqcp use}     \\
                      & d_i\in\mathbb{Z},~1\leq d_i\leq T_i                       &  & \text{for } i=1,\dots,n\label{miqcp d}       \\
                      & P_j\in\mathbb{Z},~1\leq P_j\leq \max\mathcal{T}           &  & \text{for } j=1,\dots,m\label{miqcp p}       \\
                      & f_j\in\mathbb{Q},~\frac{1}{\max\mathcal{T}}\leq f_j\leq 1 &  & \text{for } j=1,\dots,m\label{miqcp f}       \\
                      & m_{i,j}\in\{0,1\}, u_j\in\{0,1\}                          &  & \makecell[r]{\text{for } i=1,\dots,n         \\j=1,\dots,m}\label{miqcp binary}
\end{align}
It should be noted that the task periods $T_1,\dots,T_n$ and the maximal task period $\max\mathcal{T}$ are constants in the model.
Furthermore, an inequality written as $x^L\leq x\leq x^U$ is a shorthand for two separate constraints $x^L\leq x$ and $x\leq x^U$.
The model introduces the following variables:
\begin{itemize}
    \item For every timer $\rho_j$ with $1\leq j\leq m$, $P_j$ denotes the period of the timer $\rho_j$ and $f_j$ denotes the expected number of interrupts that the timer causes. The timer period $P_j$ is an integer and can assume values between one and the maximal task period of $\mathbb{T}$. This is because if a timer is configured, it must have an interrupt interval of at least one time unit. Also, an interrupt interval is naturally bounded by the greatest task period. In the extreme case, the timer period can be equal to the maximal task period. These bounds are stated in \Cref{miqcp p}.

          As the expected number of interrupts caused by a timer $\rho_j$ is calculated as $\frac{1}{P_j}$, the range of $f_j$ is directly given by the bounds of $P_j$. The bounds of $f_j$ are defined in \Cref{miqcp f}. It should be noted that because $f_j$ is effectively calculated as $\frac{1}{P_j}$, $f_j$ is rational.

          Additionally, a binary variable $u_j$ is introduced for each timer $\rho_j$, which indicates whether $\rho_j$ should be used in the system. If $u_j=1$, timer $\rho_j$ will be used.
    \item The binary variables $m_{i,j}$ are defined for every task $\tau_1,\dots,\tau_n$ and for every timer $\rho_1,\dots,\rho_m$. A value of $m_{i,j}=1$ indicates that the task $\tau_i$ is assigned to timer $\rho_j$.
    \item For every task $\tau_i\in\mathbb{T}$, an integer variable $d_i$ is introduced. If a task $\tau_i$ is assigned to timer $\rho_j$, the timer period $P_j$ must be an integer divisor of the task period $T_i$, i.e., a number $d_i\in\mathbb{Z}$ must exist such that $T_i=d_i\cdot P_j$ holds. The bounds of $d_i$ can directly be derived from the bounds of $P_j$. Each variable $d_i$ has a lower bound of one and an upper bound of $T_i$. This is stated in \Cref{miqcp d}.
\end{itemize}
The presented MIQCP model provides a realization of the problem in the following manner:

The objective function given in \Cref{miqcp obj} minimizes the total expected number of tick interrupts.
\Cref{miqcp obj} realizes the remark about \Cref{ilp: objective function} that not necessarily every timer should be used by integrating the binary variables $u_j$ into the objective function.
Conveniently, this means that for $u_j=0$, the number of interrupts caused by timer $\rho_j$ is also zero, i.e., the timer $\rho_j$ will not be used.

The constraint given by \Cref{miqcp num} effectively assigns $f_j$ the number of interrupts caused by timer $\rho_j$, as $f_j\cdot P_j=1\Leftrightarrow f_j=\frac{1}{P_j}$.
Because \Cref{miqcp num} is a quadratic equality constraint, the model becomes non-convex~\cite{Boyd_Vandenberghe_2004}.

Next, any task $\tau_i$ must be assigned to exactly one timer $\rho_j$.
\Cref{miqcp once} ensures that this requirement is satisfied by enforcing that the sum of all variables $m_{i,j}$ for a particular task $\tau_i$ is one.
Thus, exactly one $m_{i,j}\in\{m_{i,j}\mid 1\leq j\leq m\}$ is allowed to be set to one.

The critical part of this model is the constraint that for each task $\tau_i$ that is assigned to a timer $\rho_j$, the timer period $P_j$ must be an integer divisor of the task period $T_i$.
This condition is given in \Cref{miqcp divisor} for every task $\tau_i\in\mathbb{T}$.
Naturally, $P_j$ only needs to be a divisor of $T_i$ if $\tau_i$ is assigned to $\rho_j$.
As the assignment of a task $\tau_i$ to a timer $\rho_j$ is indicated by the binary variable $m_{i,j}$ and \Cref{miqcp once} states that a task is always assigned to exactly one timer, the timer period that has to be a divisor of $T_i$ is selected by $\sum^m_{j=1}P_j\cdot m_{i,j}$.
The result of the sum is the period of the timer that task $\tau_i$ is assigned to.
If an integer $d_i$ exists that satisfies the equality constraint, $P_j$ is an integer divisor of $T_j$.
Here, it should be noted that multilinear terms like $x\cdot y\cdot z$ are often not allowed.
However, they can be rewritten by introducing auxiliary variables.
Hence, for the readability of the model, the term is left in its trilinear form.

Finally, the constraint given in \Cref{miqcp use} ensures that any binary variable $u_j$ is set to one if any task is assigned to timer $\rho_j$, i.e., if any $m_{i,j}=1$ for all tasks $\tau_i\in\mathbb{T}$.

Notably, the presented non-convex MIQCP model can be transformed into an ILP model by linearizing the quadratic constraints and objective function and reformulating the objective to remove the rational variables.
That is possible because all variables have a defined upper and lower bound.

\section{Evaluation}\label{section:Evaluation}
We evaluate the methods presented in \Cref{section:Multi-Timer} on a real-world embedded system that contains four hardware timers.
In order to evaluate the efficiency of the methods, the cumulative time spent in tick interrupts and the cumulative time that is spent on delaying tasks will be measured for every method.
The time spent on the insertion of delayed tasks is also recorded to evaluate the efficiency of \textit{Chronos-const} and \textit{Chronos-harmonic} as both methods aim to reduce the overhead caused by inserting the tasks in a sorted manner.
These measurements will then be compared to the baseline implementation in FreeRTOS~\cite{FreeRTOS}, which uses a single timer.

\paragraph{Evaluation Setup}\label{sec: eval}
\begin{figure}
    \centering
    \begin{tikzpicture}
        [rect/.style={ rectangle,
                    draw,
                    thick,
                    shape aspect=2,
                    inner sep = 5pt,
                    text centered,
                    minimum width = 1cm,
                    minimum height = .6cm,
                },
            rhom/.style={ diamond,
                    draw,
                    thick,
                    aspect=3,
                    text centered,
                    align=center
                }]
        \node[rect] (c) at (0,1.5) {Create Tasks};
        \node[rect] (start) at (0,0.5) {Start Scheduler};
        \node[rhom] (taskdef) at (0,-1.5) [align=center] {User Tasks\\ready?};

        \node[rect,anchor=east] (usertask) at (-1.25,-3) {Idle Task};
        \node[rect,anchor=west] (idle) at (1.25,-3) {User Task};

        \node[rect,fit=(usertask)(idle)(taskdef),inner sep=10pt,dashed] (running) {};

        \path[-Latex,thick] (idle) edge node [above,align=center] {\small all blocked} (usertask);
        \draw[-Latex,thick] ([xshift=0em]taskdef.west) -- ([xshift=0em]taskdef.west |- usertask.north);
        \draw[-Latex,thick] ([xshift=0em]taskdef.east) -- ([xshift=0em]taskdef.east |- idle.north);
        \node[anchor=east] at ([shift={(-.5,.45)}]usertask.north east) {\small no};
        \node[anchor=west]  at ([shift={(.5,.45)}]idle.north west) {\small yes};

        \node[rect,align=center] (system) at (0,-5) {System Tick\\Interrupt};
        \node[rect,align=center] (end) at (-3,-5) {End Scheduler};
        \node[rect,align=center] (out) at (-3,-6) {Output};
        \path[-Latex,thick] (end) edge (out);

        \path[-Latex,thick] (running) edge node [right,align=center] {\small @tick interval} (running.south |- system.north);
        \path[-Latex,thick] (start) edge  (start.south |- taskdef.north);
        \path[-Latex,thick] (c) edge  (start);
        \draw[-Latex,thick] (running.south -| end.north) -- (end.north) node [right,pos=.4] {\small all tasks done};

        \draw[-Latex,thick] (system.east) -| ([xshift=-1em] running.south east);
        \node[align=center] at ([shift={(1.1,-.75)}]system.east) {\scriptsize increment tick,\\\scriptsize release tasks,\\\scriptsize scheduling decision};

        \node[anchor=south east] (runningtask) at (running.north east) {\scriptsize Running Task};
    \end{tikzpicture}
    \caption{Structure of the FreeRTOS program}\label{fig:FreeRTOS Workflow}
\end{figure}
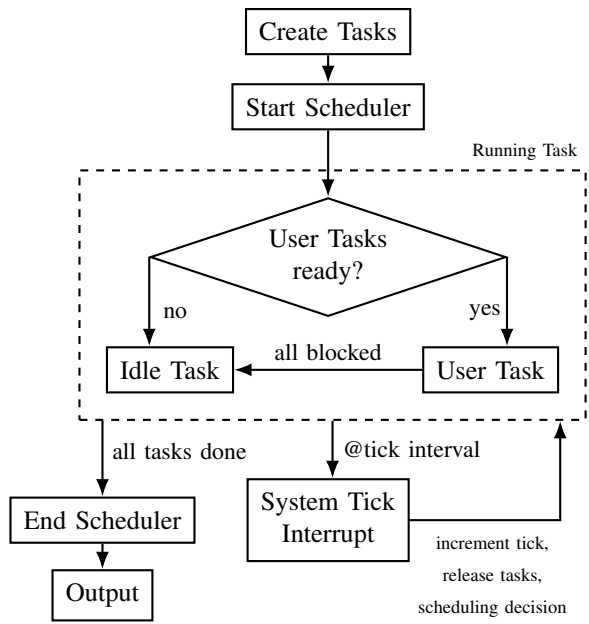
The embedded system that was used to perform the measurements is an ESP32-S3-DevKitC that contains an ESP32-S3 chip~\cite{ESP32S3} with additionally \SI{8}{\mega\byte} of externally connected PSRAM.
The ESP32-S3 is a microcontroller that contains two XTensa LX7 CPU cores that can be configured to run at up to \SI{240}{\mega\hertz}.
Internally, the chip can access \SI{512}{\kilo\byte} of SRAM.
Additionally, the chip contains four hardware timers that can be configured to trigger interrupts at periodic intervals independently of each other.
In the following, the four hardware timers will be described as $\rho_1,\rho_2,\rho_3$, and $\rho_4$ and their periods will be described as $P_1,P_2,P_3$, and $P_4$.

The methods presented in \Cref{section:Multi-Timer} were implemented in the ESP32 FreeRTOS port using the unicore version of FreeRTOS 10.5.1 included in the ESP-IDF 5.2.1 development framework.
Every test was performed with enabled preemption and time slicing.
The included \texttt{GPTimer} library was used to configure the hardware timers.

For all tests, the CPU frequency was set to \SI{240}{\mega\hertz} without dynamic frequency scaling.
For the baseline implementation in FreeRTOS, the system timer period was always set to \SI{1}{\milli\second}.
The additional hardware timers were configured such that one tick in the multi-timer setup also corresponds to \SI{1}{\milli\second}.
For this, the frequency at which the internal counters of the hardware timers operate was configured to \SI{1}{\mega\hertz}.
Then, for a timer $\rho_j$ with a period of $P_j$ ticks, the timer was configured to generate an interrupt when the internal counter reaches a value of $P_j\cdot 1000$, i.e., an interrupt is generated after $P_j\cdot\SI{1}{\milli\second}$.
When an interrupt is triggered, the counter is always reset to zero and starts anew.

\paragraph{Structure of the FreeRTOS program}
In the following, the structure of the program that is used to evaluate the methods is outlined.
A visualization of the structure of the program can be found in \Cref{fig:FreeRTOS Workflow}.
Generally, in FreeRTOS, it is allowed to create tasks after the scheduler was started, i.e., the task set can grow dynamically.
However, as the task set needs to be known in advance to create the mapping of tasks to timers, the program takes a predefined task set as an input, and all tasks are created before the FreeRTOS scheduler is started.
After the scheduler is started, the execution of the created tasks starts.
Every task is configured with a fixed workload and a fixed number of releases.
When all tasks have finished their workload, the scheduler is ended and the collected information, such as the cumulative time spent in tick interrupts, total runtime and number of deadline misses, is produced as an output.

\paragraph{Measurement Methodology}\label{Evaluation::Setup::Methodology}
In order to evaluate the overhead caused by the different methods, the overall time spent in tick interrupts has to be measured.
For this, probe points are added at the start and the end of any interrupt to record the time spent in that interrupt.

To differentiate tick interrupts from other interrupts, such as yield calls that are also implemented via interrupts in the ESP32 port of FreeRTOS, the source of the interrupt is inspected before any measurements are taken.
As the ESP32-S3 is based on the XTensa architecture~\cite{xtensa}, there are 32 interrupt sources that can be allocated to different interrupt handlers.
Every interrupt is identified in the system by a number between 0 and 31, which is assigned to that interrupt when it is allocated.
When an interrupt occurs, the \texttt{INTERRUPT} register can be read to check which interrupt was triggered.

The \texttt{INTERRUPT} register is a 32-bit register where the $i$-th bit represents the interrupt that was allocated the number $i$.
That means, if bit $k$ of the \texttt{INTERRUPT} register is one when an interrupt occurs, the interrupt source that was assigned the number $k$ triggered the interrupt.
In order to only measure the time spent in tick interrupts, the mapping of interrupt sources to bits of the \texttt{INTERRUPT} register was fixed.
Then, at the start of each interrupt, it is checked whether the bit that represents a tick interrupt source is set to one.
If that is the case, a timestamp in the form of the current clock cycle is recorded.
For this, the \texttt{CCOUNT} register can be read.
At the end of an interrupt, another timestamp is recorded, if a timestamp was also recorded at the start of the interrupt.
Additionally, to measure the time that is spent on delaying tasks, probe points are added to the \texttt{xTaskDelayUntil()} and \texttt{delay()} methods.

\paragraph{Test Configuration}\label{Evaluation::Setup::Test Configuration}
To evaluate the different methods, several test scenarios were formed.
First, \textit{Chronos} and \textit{Chronos-const} were evaluated against the baseline FreeRTOS implementation using a non-harmonic task set.
For this, two different workload scenarios were evaluated where the amount of work that each task has to perform differs.
In the \textit{low} single-job workload scenario, each task has to perform 1000 additions before it will be delayed again.
For the \textit{high} single-job workload scenario, each task performs 10000 additions during its execution.
This was done to explore how the overhead of the tick interrupts changes when tasks are more likely not to be delayed already when the tick interrupt occurs next.
A higher single-job workload means that each task needs more time to execute and thus spends more time in the \texttt{ReadyList}.
Thus, the number of tasks that need to be processed during a tick interrupt tends to be less, resulting in a shorter interrupt duration.

Generally, each task is configured to be released five times before it is considered to be finished.
After five releases, the task will be blocked immediately after it has been released.
Furthermore, the methods are also evaluated for harmonic task sets, such that \textit{Chronos-harmonic} can also be applied.
Here, the same low and high single-job workloads are employed, albeit for tasks with different periods, as the task set is harmonic.

Additionally, a scenario called \textit{harmonic single} is evaluated in which the baseline implementation is compared to a single timer setup using the techniques from \textit{Chronos-const} and \textit{Chronos-harmonic} with a fixed timer period of \SI{1}{\milli\second}.
This scenario compares the efficiency of the different methods without reducing the number of interrupts.
However, in this scenario the number of additions each task has to perform was reduced to 100, such that no deadline misses occur.

Since the ESP32-S3-DevKitC is an embedded system with only moderate amounts of memory and computational power, it cannot support very large task sets without experiencing deadline misses.
For this reason, both the non-harmonic and harmonic task sets contain 100 tasks.
For the multi-timer methods, the MIQCP model from \Cref{section:Task Distribution} was used to distribute the tasks to the four timers and to configure the periods of the timers.
The result is a partition $\Gamma=\{\mathbb{T}_1,\mathbb{T}_2,\mathbb{T}_3,\mathbb{T}_4\}$ of the task set $\mathbb{T}$ where timer $\rho_j$ is assigned the tasks of the set $\mathbb{T}_j$, for $1\leq j\leq 4$.
For both the non-harmonic and the harmonic task sets, the tasks were generated with periods that are multiples of $\SI{3}{\milli\second},\SI{5}{\milli\second},\SI{7}{\milli\second}$, or $\SI{11}{\milli\second}$.
Therefore, the four timers were also configured to generate interrupts at $\SI{3}{\milli\second},\SI{5}{\milli\second},\SI{7}{\milli\second}$, or $\SI{11}{\milli\second}$.
For the \textit{harmonic single} test, only a single timer with a period of \SI{1}{\milli\second} was used.
In the non-harmonic task set, for every task a base period $b\in\{\SI{3}{\milli\second},\SI{5}{\milli\second},\SI{7}{\milli\second},\SI{11}{\milli\second}\}$ was chosen and then multiplied with a random factor $r\in\{x\in\mathbb{N}\mid 1\leq x\leq 10\}$, such that for a timer $\rho_j$, all its associated tasks have a period of $T_i=P_j\cdot r$, for all $\tau_i\in\mathbb{T}_j$ and $P_j\in\{\SI{3}{\milli\second},\SI{5}{\milli\second},\SI{7}{\milli\second},\SI{11}{\milli\second}\}$.
For the harmonic task sets, the same approach was used, but the random factor $r$ was always selected from the set $\{2^i\mid i\in\{0,1,2,3,4\}\}$ to only create tasks with harmonic periods.
The random factor $r$ was always sampled from a uniform distribution.
We used Gurobi~\cite{gurobi} to implement and solve the MIQCP.

The generated tasks have periods that are multiples of the timer periods to be more representative of realistic test scenarios.
If all tasks had the same period as their associated timer, its \texttt{TimerList} would always contain all tasks of that partition when the tick interrupt occurs, assuming no task misses its deadline.
Consequently, all tasks would always need to be released then as well.
A summary of the different test configurations can be found in \Cref{configtable}.

Finally, to evaluate the impact of the different methods for task sets with increasing task periods, a common \textit{period factor} $p\in\mathbb{N}$ was introduced that is used to uniformly scale the timer and task periods.
For a given period factor $p$, all timer periods and all task periods were multiplied by $p$.
Every scenario was evaluated for period factors between 1 and 15, as testing has shown that the trends continue in the same manner for larger period factors.

\begin{table}
    \centering
    \begin{tabular}{l@{\hspace{.5cm}}rrr }
        \toprule
        Workload        & Number of additions & Timer periods \\\midrule
        low             & 1000                & $3,5,7,11$    \\
        high            & 10000               & $3,5,7,11$    \\\midrule
        harmonic single & 100                 & 1             \\
        harmonic low    & 1000                & $3,5,7,11$    \\
        harmonic high   & 10000               & $3,5,7,11$
        \\\bottomrule
    \end{tabular}\vspace{1ex}%
    \caption{Overview of all test configurations}\label{configtable}
\end{table}

\subsection{Non-Harmonic Task Sets}\label{sec: eval nonharmonic}
\begin{figure}
    \centering
    \includegraphics{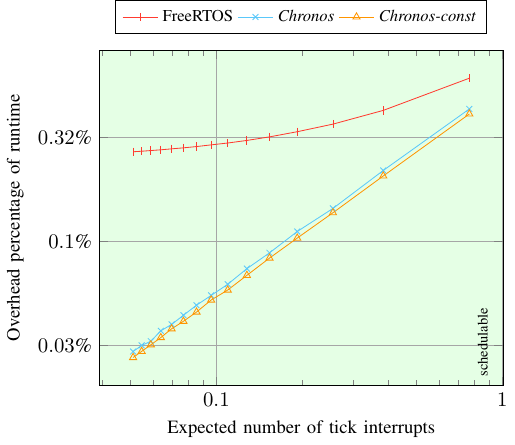}
    \caption{Non-harmonic task sets with a low single-job workload}\label{nonharmonic-low}
\end{figure}
In the following, \textit{Chronos} and \textit{Chronos-const} are evaluated against the baseline FreeRTOS implementation that uses a single timer.

In every plot, the period factors for which the task set experiences deadline misses regardless of the task dispatching method is colored with a red background and labeled \textit{not schedulable}.
For period factors where no deadline misses occur with any method, the background is colored green and labeled \textit{schedulable}.
In general, a task set is called \textit{schedulable} here, if under any method, the task set can be executed without any deadline misses.
If, for a particular period factor, deadline misses occur when using the baseline implementation but not when using \textit{Chronos} or \textit{Chronos-const}, the region is colored with a yellow background and labeled \textit{Chronos}.
Lastly, if all methods other than \textit{Chronos-harmonic} experience deadline misses, the region is colored with a blue background and labeled \textit{harmonic}.
It should be noted that the range of the y-axis differs between plots of different scenarios as the range of the measured data also differs vastly.

\paragraph{Low Workload Scenario}\label{sec: eval nonharmonic low}
First, the overhead measurements for non-harmonic task sets with a low single-job workload are presented in \Cref{nonharmonic-low}.

The y-axis shows the percentage of the total runtime that is spent on delaying tasks and handling tick interrupts.
On the x-axis, the expected number of tick interrupts when using \textit{Chronos} or \textit{Chronos-const} is shown for period factors $p\in[1,15]$.
This is normalized to the expected number of tick interrupts when using the baseline implementation and can be calculated as $\frac{1}{P_1}+\frac{1}{P_2}+\frac{1}{P_3}+\frac{1}{P_4}$.
A value of $0.5$ as the expected number of tick interrupts means that the multi-timer methods cause half the number of tick interrupts compared to the baseline in a fixed time interval.
It should be noted that both axes of the overhead measurements are logarithmically scaled since the values on both axes are fractions.

\begin{figure}
    \centering
    \includegraphics{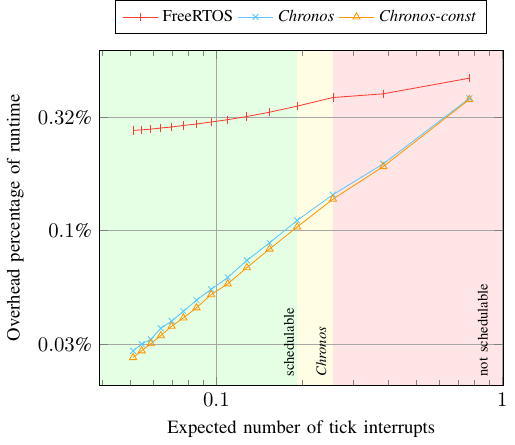}
    \caption{Non-harmonic task sets with a high single-job workload}\label{nonharmonic-high}
\end{figure}
At first glance, it can be seen that the task sets for all period factors are schedulable under all task dispatching methods.
The measurements show that \textit{Chronos} and \textit{Chronos-const} provide significant reductions in overhead compared to the baseline implementation in FreeRTOS that only uses a single timer.
Furthermore, for increasing period factors, the percentage of the runtime that is spent on blocking tasks and handling tick interrupts constantly decreases.
The overhead for the baseline implementation in FreeRTOS also decreases, albeit considerably slower.
This can be explained by the number of tick interrupts that occur when using one of the multi-timer methods compared to the baseline implementation.
For \textit{Chronos} and \textit{Chronos-const}, the number of interrupts that occur does not change for longer period factors.
This is because all task periods are uniformly scaled with the period factor, resulting in the GCD of every partition also being scaled in the same manner.
Also, for longer running programs, the percentage of the overall runtime that is spent on delaying tasks and handling tick interrupts also naturally decreases as only a small portion of the runtime is overhead.

Another observation can be made about the difference in overhead caused by \textit{Chronos} and \textit{Chronos-const}.
While both methods cause a similar amount of overhead, \textit{Chronos-const} always causes slightly less overhead compared to \textit{Chronos}.
In this scenario, the trade-off made in \textit{Chronos-const} is worthwhile with regard to the overhead.
All in all, for this task set, the multi-timer methods provide a considerable reduction in overhead compared to the baseline implementation for all period factors.

\paragraph{High Workload Scenario}
In \Cref{nonharmonic-high}, the measurements for non-harmonic task sets with a high single-job workload are presented.
At first glance, it can be seen that for period factors of one to three, deadline misses occur regardless of the employed task dispatching method.
Also, for a period factor greater than three, only when using \textit{Chronos} or \textit{Chronos-const} no deadline misses occur.
That means, by changing the dispatching method, it is possible to make a task set schedulable that was previously not.
Finally, for period factors greater than four, the task set becomes schedulable under all methods.

\begin{figure}
    \centering
    \includegraphics{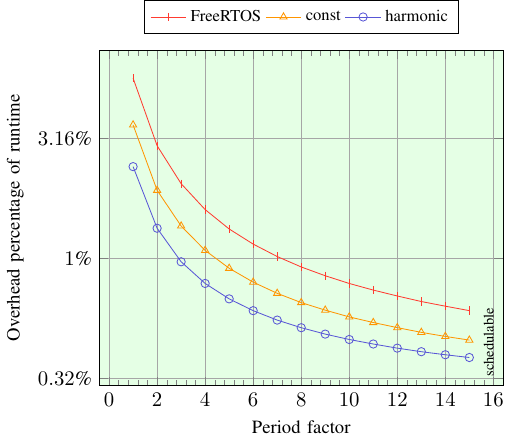}
    \caption{Harmonic task sets with a single timer}\label{harmonic-ST}
\end{figure}
The trends of the overhead measurements match those of the low workload scenario.
However, it can be observed that for a period factor of one, the overhead caused by \textit{Chronos} is only slightly higher than the overhead caused by \textit{Chronos-const}, meaning that for task sets where many deadline misses occur, \textit{Chronos} can provide a similar system efficiency to \textit{Chronos-const}.
Nevertheless, \textit{Chronos-const} still performs slightly better for all period factors.

\begin{figure}
    \centering
    \includegraphics{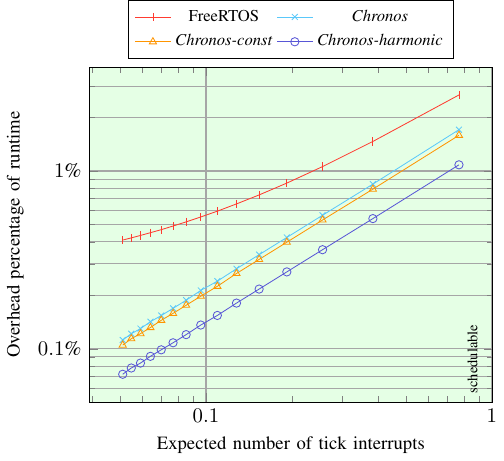}
    \caption{Harmonic task sets with a low single-job workload}\label{harmonic-low}
\end{figure}

\subsection{Harmonic Task Sets}\label{sec: eval harmonic}
We further conducted experiments where the examined task sets are always harmonic.
Therefore, it is possible to also evaluate \textit{Chronos-harmonic} alongside \textit{Chronos} and \textit{Chronos-const} against the baseline implementation.

\paragraph{Single Timer Tests}
First, in \Cref{harmonic-ST}, the impact on the overhead by using the techniques from \textit{Chronos-const} and \textit{Chronos-harmonic} without using multiple timers or adapting the timer period to the GCD of the task set are presented.
Concretely, the data points for \textit{const} show the effect of sacrificing an ordered \texttt{TimerList} for a faster task insertion.
The data points for \textit{harmonic} show the effects of using the fixed-array-based dispatching method from \Cref{sec:Chronos-harmonic}.
For this test, all timer periods were fixed at \SI{1}{\milli\second}, and the number of additions a task has to perform was set to 100.
It should be noted that, here, the period factor is shown on the x-axis as all methods use a single timer that has a fixed period.

At first glance, it can be seen that regardless of the period factor, all task sets are schedulable.
While both methods improve the system efficiency in comparison to the baseline, the \textit{harmonic} method further reduces the overhead compared to the \textit{const} method.
For this task set and test configuration, it is beneficial to utilize the \textit{const} or \textit{harmonic} methods even when only a single timer can be used whose period cannot be extended.

\paragraph{Low Workload Scenario}\label{sec: eval: harmonic-low}
In \Cref{harmonic-low}, the measurements for harmonic task sets with a low single-job workload are shown.
Generally, all multi-timer methods provide a significant reduction of the overhead compared to the baseline.
While \textit{Chronos} and \textit{Chronos-const} show similar results, using \textit{Chronos-harmonic} can further reduce the overhead.
In this scenario, the overhead decreases faster for increasing period factors for the multi-timer methods compared to the baseline implementation.

\paragraph{High Workload Scenario}\label{sec: eval: harmonic-high}
\begin{figure}
    \centering
    \includegraphics{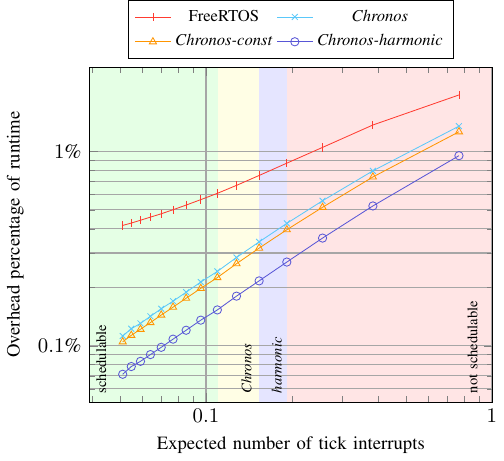}
    \caption{Harmonic task sets with a high single-job workload}\label{harmonic-high}
\end{figure}

Next, the measurements for the high single workload scenario for harmonic task sets are given in \Cref{harmonic-high}.
This plot shows that using \textit{Chronos-harmonic} can enable more task sets to be schedulable compared to using \textit{Chronos} or \textit{Chronos-const}.
In more detail, for period factors up to four, the task sets are not schedulable under any task dispatching method, while for period factors greater than four, task sets are schedulable when using \textit{Chronos-harmonic}.
Notably, only for period factors greater than five, task sets become schedulable under \textit{Chronos} and \textit{Chronos-const}.
When using the baseline implementation, task sets become schedulable for period factors greater than seven.
As a result, here, using \textit{Chronos-harmonic} instead of \textit{Chronos} or \textit{Chronos-const} can allow further task sets to be schedulable.

\subsection{Discussion}\label{sec: eval discussion}
In this section, the methods presented in \Cref{section:Multi-Timer} were evaluated on an embedded system.
For all tested configurations, it can be observed that utilizing multiple hardware timers can reduce the overall time overhead.
Also, for scenarios where a task set is not schedulable under the default FreeRTOS implementation with a single timer, it is possible that when employing one of the multi-timer methods, that task set will be schedulable.
Additionally, for harmonic task sets, it can be beneficial to exploit the way harmonic tasks are released to further reduce the overhead.
Moreover, for scenarios where only a single timer can be used whose period cannot be extended, the \textit{const} and \textit{harmonic} methods can still reduce the overhead compared to the baseline implementation.

The evaluation results are summarized in \Cref{summarytable}.
For every tested scenario, the peak reduction of the overhead when using one of the presented methods compared to the baseline implementation is listed.
In addition, the geometric mean of the overhead reduction over all task sets that are schedulable under any method is also listed for every presented method.
The results for \textit{Chronos} are given in the columns labeled \textit{chronos}, and the results for \textit{Chronos-const} are shown in the columns labeled \textit{const}.
For the harmonic task sets, the results for \textit{Chronos-harmonic} are listed in the columns labeled \textit{harmonic}.

For the tests with non-harmonic task sets, \textit{Chronos-const} achieves a reduction of the overhead of $9.79\times$--$10.01\times$ in peak and $4.94\times$--$6.37\times$ on average.
\textit{Chronos} is slightly less efficient and reduces the overhead by $9.16\times$--$9.36\times$ in peak and $4.64\times$--$5.98\times$ on average.
For the harmonic task sets with only a single timer, the \textit{const} method reduces the overhead by $1.56\times$ in peak and $1.41\times$ on average.
The \textit{harmonic} method is more efficient and reduces the overhead by $2.33\times$ in peak and $1.83\times$ on average.
Next, for scenarios with multiple timers, \textit{Chronos} reduces the overhead by $3.64\times$--$3.7\times$ in peak and $2.52\times$--$3.07\times$ on average, while \textit{Chronos-const} achieves an overhead reduction of $3.89\times$--$3.97\times$ in peak and $2.69\times$--$3.3\times$ on average.
\textit{Chronos-harmonic} is more efficient here as it reduces the overhead by $5.68\times$--$5.83\times$ in peak and $3.94\times$--$4.82\times$ on average.

\begin{table}
    \centering
    \resizebox{\linewidth}{!}{%
        \begin{tabular}{lcccccc}
            \toprule
            Scenario & \makecell{Peak                                                                             \\\textit{chronos}}    & \makecell{Mean\\\textit{chronos}}    & \makecell{Peak\\\textit{const}} & \makecell{Mean\\\textit{const}} & \makecell{Peak\\\textit{harmonic}} & \makecell{Mean\\\textit{harmonic}} \\\midrule
            Non-harmonic                                                                                          \\
            low      & $9.16\times$   & $4.64\times$ & $9.79\times$  & $4.94\times$ & ---          & ---          \\
            high     & $9.36\times$   & $5.98\times$ & $10.01\times$ & $6.37\times$ & ---          & ---          \\\midrule
            Harmonic                                                                                              \\
            single   & ---            & ---          & $1.56\times$  & $1.41\times$ & $2.33\times$ & $1.83\times$ \\
            low      & $3.64\times$   & $2.52\times$ & $3.89\times$  & $2.69\times$ & $5.68\times$ & $3.94\times$ \\
            high     & $3.7\times$    & $3.07\times$ & $3.97\times$  & $3.3\times$  & $5.83\times$ & $4.82\times$ \\\bottomrule
        \end{tabular}}\vspace{1ex}%
    \caption{Reduction of overhead compared to FreeRTOS}\label{summarytable}
\end{table}

\section{Conclusion}\label{chapter:Conclusion}
In this paper, we examined how to utilize multiple hardware timers to reduce the overhead of the task handling process in an RTOS.
We formulated an MIQCP that partitions a given task set and assigns each partition to a different hardware timer, such that the overall number of tick interrupts is minimized.
Additionally, we also investigated how to further reduce the overhead by changing the way the set of delayed tasks is organized.
The first method, \textit{Chronos}, distributes the task set over multiple timers and keeps a sorted list for the delayed set of every timer.
\textit{Chronos-const} shifts the organization overhead from the task to the interrupt handler, by keeping an unstructured list for the set of delayed tasks that has to be iterated completely in every tick interrupt.
Lastly, \textit{Chronos-harmonic} is an optimization specifically for harmonic task sets.
Here, the properties of harmonic tasks are exploited to enable a constant time insertion and retrieval of delayed tasks.

In order to examine the impact on the overall time overhead caused by the different methods, multiple scenarios for harmonic and non-harmonic task sets were evaluated on a real-world embedded system.
For non-harmonic task sets, both \textit{Chronos} and \textit{Chronos-const} are more efficient than the baseline implementation of FreeRTOS with a single timer.
Generally, for the tested configurations, \textit{Chronos-const} always outperforms \textit{Chronos} and reduces the overhead compared to the baseline by up to $\approx10\times$ in peak and $\approx6\times$ on average.

As \textit{Chronos-harmonic} can only be applied to harmonic task sets, several tests were performed with harmonic task sets.
First, it was evaluated whether the techniques used by \textit{Chronos-const} and \textit{Chronos-harmonic} also reduce the overhead when only a single timer with a fixed period is used.
In this scenario, using \textit{Chronos-harmonic} results in the least overhead, as the overhead is reduced by up to $2.33\times$ in peak and $1.83\times$ on average compared to the baseline.
For scenarios where multiple timers could be used, \textit{Chronos-harmonic} performs the best by reducing the overhead by up to $5.83\times$ in peak and $4.82\times$ on average.

In conclusion, the presented methods can be used to reduce the overhead in the task handling process that is incurred by an RTOS, if the hardware platform provides multiple timers. As a result, task sets that were not schedulable when a single timer is employed, can be schedulable with multiple timers.

\bibliographystyle{IEEEtran}
\bibliography{IEEEabrv,literature}

\end{document}